# OWPCP: A Deep Learning Model to Predict Octanol-Water Partition Coefficient

*Mohammadjavad Maleki[*1], Sobhan Zahiri[†2]*

1. Department of Chemistry, Sharif University of Technology, Azadi Ave, Tehran, Iran

2. Department of Chemistry, Isfahan University of Technology, University Boulevard, Esteghlal Square, Isfahan, Iran

ABSTRACT

The physicochemical properties of chemical compounds have great importance in several areas, including pharmaceuticals, environmental science, and separation science. Among these are physicochemical properties such as the octanol-water partition coefficient log P, which has been considered an important index pointing out lipophilicity and hydrophilicity. It affects drug absorption and membrane permeability. Following Lipinski's rule of five, log P was identified as one of the key determinants of the stability of chemical entities and, as such, needed state-of-the-art methods for measuring lipophilicity. This paper presents a deep-learning model named OWPCP, developed to compute log P using Morgan fingerprints and MACCS keys as input features. It uses the interconnection of such molecular representations with log P values extracted from 26,254 compounds. The dataset was prepared with care to contain a wide range of chemical structures with differing molecular weights and polar surface area. Hyperparameter optimization was conducted using the Keras Tuner alongside the Hyperband algorithm to enhance the model's performance. OWPCP demonstrated outstanding predictive performance compared to current computational methods, achieving an MAE of 0.247 on the test set and outperforming all previous deep learning models. Most remarkably, while one of the most accurate recent models is based on experimental data on retention time to make predictions, OWPCP manages computing log P efficiently without depending on these factors, being, therefore, very useful during early-stage drug discovery. Also, our model outperforms the best model, which leverages Retention Time, and our model does not require any experimental feature. Further validation of the model performance was done across different functional groups, and it showed very high accuracy, especially for compounds that contain aliphatic hydroxy groups. The results have indicated that OWPCP provides a reliable prediction of log P.

INTRODUCTION

Physicochemical properties can be categorized into two subgroups: molecular properties and bulk properties. The former subgroup includes properties such as molecular weight, dipole moment, polarizability, van der Waals volume, and surface area, and the latter subgroup contains properties such as acidic or basic character in solution, octanol-water partition coefficient, solubility, etc. [1]. Among these properties, the octanol-water partition coefficient has vast applications in separation science and other important fields of science, such as industrial, environmental, and especially pharmaceutical studies [1-3]. The most common way to express the lipophilicity and hydrophilicity of a chemical compound is the octanol-water Partition Coefficient. This notation, log P, is a numerical representation of a compound distribution between n-octanol and water [2], [3]. In other words, a molecule with a low log P value indicates mostly polar character and vice versa [1]. According to Lipinski's Rule of Five, octanol-water partition coefficients below 5 are acceptable for drug candidates [4], [5]. In addition, it has been reported in the literature that a potential drug should have a log P between 1 and 3 [6], [7]. Values below 1 imply poor membrane permeability, while values above 3 demonstrate low absorption [8]. Given the importance of log P, various experimental approaches have been developed for the measurement of log P [9-11], including the shake flask method and reversed-phase high-performance liquid chromatography (RPLC) [12], [13]. Nonetheless, determining log P by experimental methods is challenging, particularly for molecules that are unstable in solvents. Furthermore, some compounds may not be commercially available, or they may be difficult to synthesize. These pitfalls of experimental methods raise concerns at the early stages of drug discovery when there is a need for the determination of the physicochemical properties of a sheer number of compounds [14]. Accordingly, demands on fast, accurate, and high-quality methods for the prediction of log P have brought attention to computational methods. In addition, the chemical structure of a molecule determines the correlation between the molecular properties and bulk properties of the molecule [1]. Hence, various computational methods have been developed to predict log P, a bulk property, employing physiochemical properties, Liquid Chromatography Retention Time, Topographic maps, Extended Connectivity Fingerprints (ECFP), and/or sometimes molecular properties [15-22]. Here, we put forward OWPCP, a deep-learning model that estimates the Octanol-water partition coefficient by using Morgan fingerprints and MACCS keys as input features. In turn, the Morgan fingerprint is one of the circular fingerprint approaches, the purpose of which is to capture molecular substructure by iterative atom encoding. Iteratively updating atom identifiers based on the neighboring atoms and bonds over multiple generations allows the capture of the local chemical environment. The resulting identifiers are processed as a binary bit array, resulting in a fingerprint useful for modeling [23], and Molecular ACCess System keys encompass the presence of unique substructures, such as Carbon-Carbon double bonds, or carbonyl functional groups, etc., within the compounds [24]. Using Morgan fingerprints along with MACCS keys to represent each chemical compound, OWPCP learns to understand the relationship between fingerprints and log P. Precisely, a representation of the Morgan fingerprints and a representation of the MACCS keys are learned concurrently. These encodings are then inputted into a multilayer perceptron deep neural network that computes the log P. A comprehensive examination of OWPCP brought to light its remarkable performance in comparison to different state-of-the-art and earlier computational techniques.

METHODS

**Dataset.** For obtaining the dataset, experimental log P values of chemical compounds and corresponding SMILES were derived from the datasets [23], [24]. Ulrich et al. [15] processed the dataset [24] and removed the data points with an erroneous nature, yielding 13,889 chemicals. This dataset was downloaded from GitHub Respiratory (https://github.com/nadinulrich/log_P_prediction). Combining datasets, which contained numerous classes of chemical compounds, and removing duplicates resulted in a dataset with 26,254 compounds with known experimental log Ps. For all the steps, 10% of the data was randomly split as the independent test set, while the remaining 90% was used for hyperparameter tuning and training the model. The remaining data was distributed into an 80% training set and a 20% validation set, which was used to train and select the best hyperparameters. Following the preparation of the dataset, RDKit and SMILES strings [25] were utilized to generate the Morgan fingerprints (nbits of 2048 and radius of 4) [26] and MACCS keys [27] as input features for the ML model. The diversity of the dataset was assessed by determining the distribution of Molecular Weights (MW) and Polarized-Surface Areas (PSA) of the compounds, which were calculated utilizing RDKit [25]. The distributions of the log P, MWs, and PSAs were plotted using Matplotlib [28].

**Machine Learning (ML) Model.** ML model was implemented in the Keras package [29], [30] of Python [31]. Multilayer perceptron, fully connected deep neural networks that leverage backpropagation to train [32], were used as the base of the model. Taken from Emad et al. [33], three distinct MLPs were connected, as depicted in Figure 1. The model architecture includes two separate MLPs, which receive the feature vectors of compounds, Morgan fingerprints, and MACCS keys vectors. Both encoders are consisted of 3 layers (input, hidden, and output). The outputs of these two parallel and separately parametrized encoders are concatenated to serve as the input for the log P predictor, another MLP. Finally, the predictor MLP learns to predict the log P based on the outputs of the encoders. Since the log P values are real numbers, the linear activation function was chosen as the activation function of the log P predictor's output layer. This output layer also consists of 1 neuron.

**Hyper-Parameter Tuning.** To idealize the hyperparameters, the Keras tuner [34], in conjunction with the Hyperband algorithm [35] within the Keras package [29], [30] were employed. The hyperparameters that were evaluated in the optimization process were the number of neurons (Units), activation functions, dropout rates, and the learning rate. The number of neurons for each layer, except for the last output layer that was kept constant, was analysed between 512 and 4096. Rectified Linear Unit (ReLU) [36], Exponential Linear Unit (ELU) [37], linear and Tanh hyperbolic tangent function [32] activation functions were evaluated. Learning rates of 1×10-2, 1×10-3, 1×10-4, 1×10-5, and 1×10-6 were assessed. Simultaneously, dropout regularizations between 0 and 0.5 were introduced to the hidden layers to reduce overfitting.

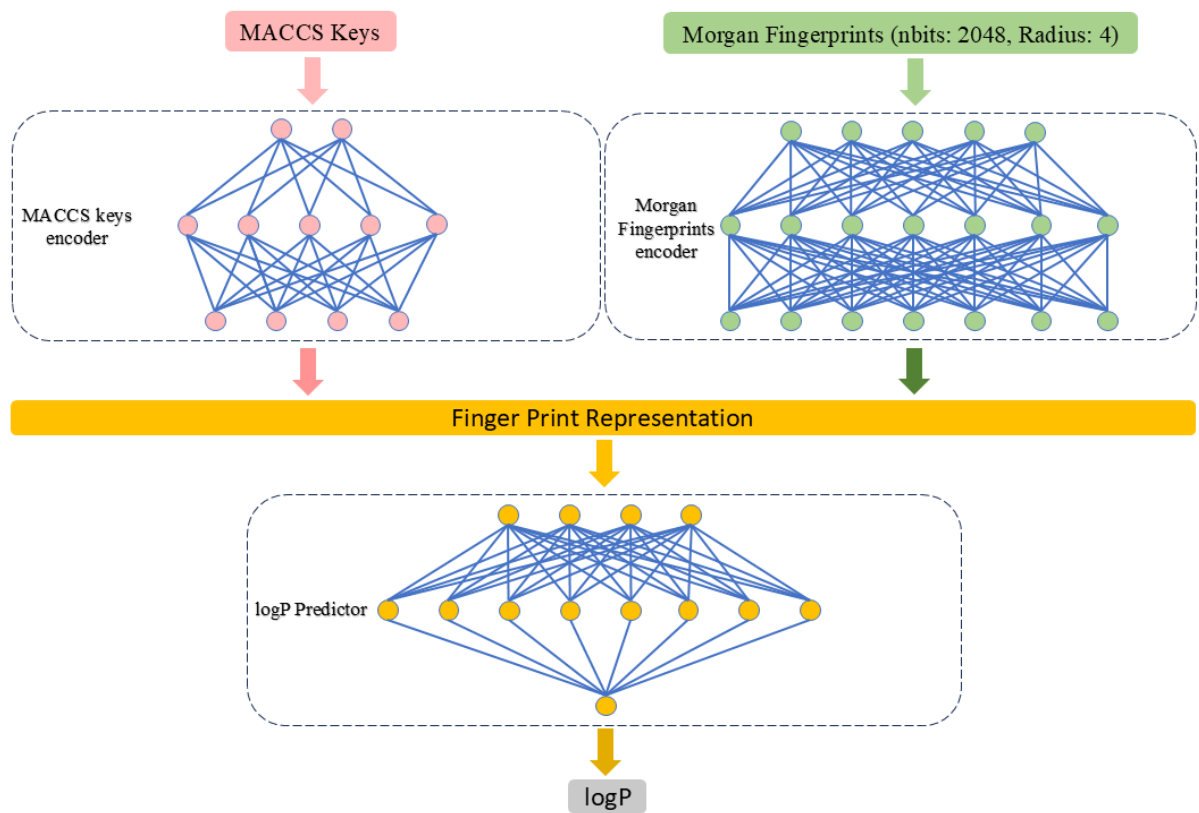

**Figure 1.** A detailed diagram of the model architecture

RESULTS AND DISCUSSIONS

**Data Diversity.** Figure 2 demonstrates the distribution of the three features of the dataset, including Molecular Weight (MW), polarized surface Area (PSA), and log P values. PSA determines the polar part of a molecule [38]. Furthermore, the membrane permeability of a drug-like molecule is affected by PSA [39]. PSA and MW are two important features affecting the drug ability and the membrane permeability of chemicals [4]. In fact, a dataset for drug design purposes and solubility predictions requires structures with a range of PSAs and MWs. Additionally, the dataset should encompass a wide range of log P values. Log P, MW, and PSA values were in the range of -5-11.5, 0-1200, and 0-300, respectively. For further study, Q-Q plots were plotted. As shown in Figure 2, Log p values of the data are almost normally distributed with a slightly positive deviation on the right tail. However, MW and PSA Q-Q plots imply positive deviations on both tails. With some estimations, data could be considered approximately normally distributed, which states the acceptable diversity of the chemical compounds within the dataset.

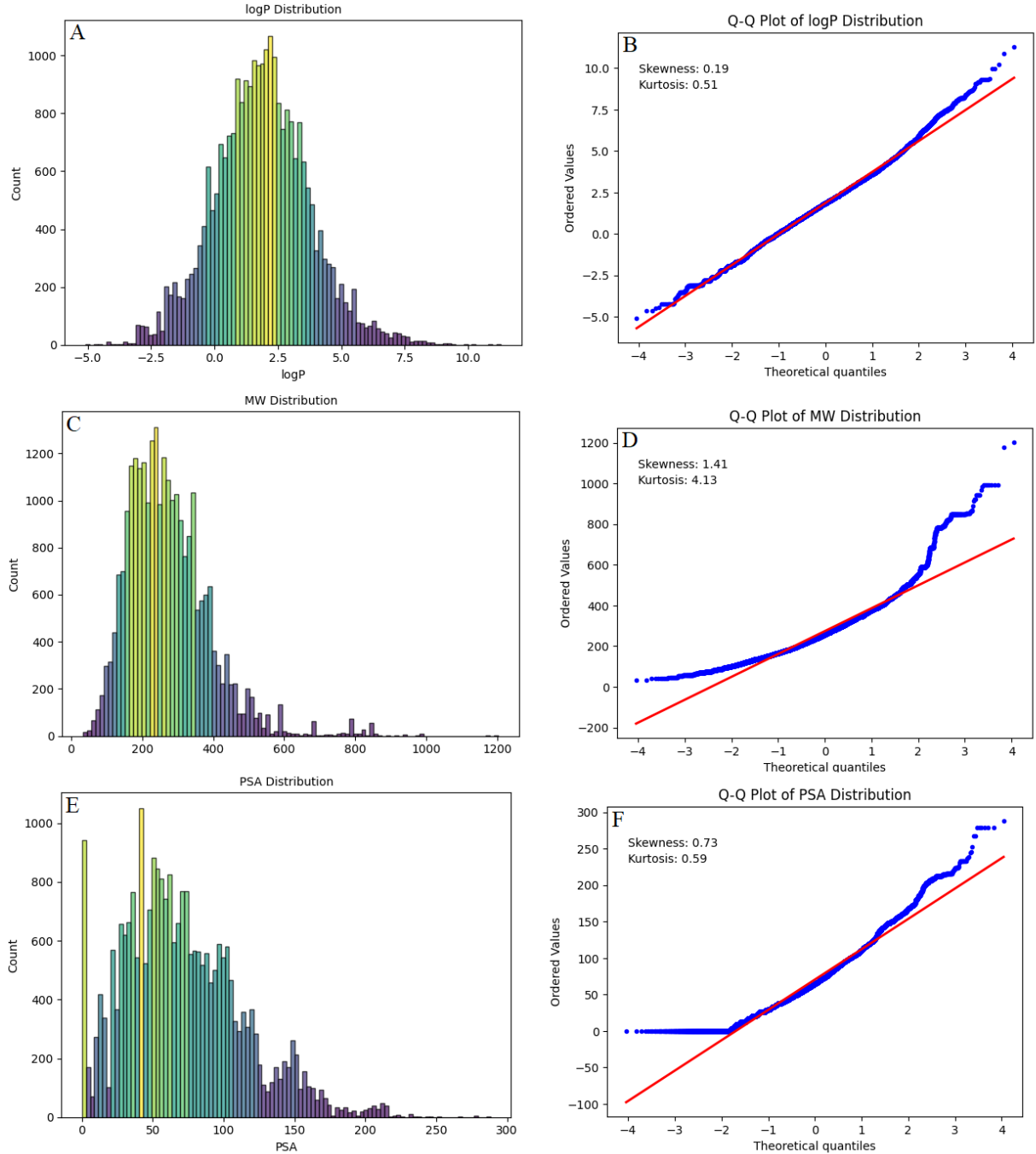

**Figure 2.** Histograms and Q-Q plots of MWs, PSA, and log P distributions of the dataset

**ML Model Development.** By the Hyperband algorithm, the hyperparameters were optimized (Table S1). Validation Mean-Squared Error was monitored during the hyperparameter tuning. The model was implemented based on the optimized hyperparameters, which resulted from the optimization step. Data were again randomly divided into test, train, and validation sets, and the model with optimized hyperparameters was trained on them, utilizing, again, the Mean-Squared Error (MSE) of the validation set to optimize the weights and biases of the model. As shown in Figure 3, MSE for both training and validation data reduced continuously over the epochs, indicating that the model was learning and minimizing the error. While the training MSE shows significant improvements in the first 40 epochs, the validation MSE does not show significant improvements after 20 epochs, revealing that the optimization process converged after 20 epochs. For the purpose of improving accuracy and reducing errors, the optimization process was carried out over a total number of 100 epochs. The trend of Mean Squared Error over the epochs is depicted in Figure 3. The best model was chosen based on the validation MSE, and thereafter it was used to predict the log p values for the data in the test set. The log p values predicted by the model were used to calculate the RMSE, MSE, MAE, and R2 (Figure 4) metrics for the test set, shown in Table 1. Test set accuracies indicated that our model was well adapted to predict the values more accurately than recently developed models. For instance, Datta et al. developed an MLP using the DeepChem database [19]. While their MLP performs well with an approximate precision of MAE = 0.477 on Log P predictions, our model showed significant improvement, attaining an MAE of 0.247 on the prediction of the test set, 48% smaller than the error that Datta et al. reported. Ulrich et al. reported a deep neural network model with RMSE = 0.47 [15]. According to the performance of our model with RMSE = 0.413, our model is more accurate than Ulrich et al. Deep Neural Network model by 12%. While Win et al. achieved the MSE of 0.26 with only a training set including roughly 2100 compounds [40], our model has shown an MSE of 0.171 but with 26,250 data points. The model reported by Win et al. achieved great precision with a very limited number of instances and performed better than our model in this regard. However, the model of Win et al. requires the Retention Time (RT) to achieve this precision. The need for an experimental value could be of significant importance when screening the library of compounds with a large number of instances, especially when the compounds are not commercially available or in cases when the chemicals to be tested are very large in numbers. In spite of the model provided by Win et al., our model does not need experimental data such as RT to predict the value of log P. Additionally, the accuracy of our model, regardless of the large used library, is superior to the model developed by Win et al. Furthermore, our model's error with an RMSE of 0.413 on the test set is notably low when compared with the experimental error, which is in the range of 0.2–0.4 log units [15].

**Table 1.** Model metrics on training, test, and validation sets

| Metric | Training | Validation | Test |
| --- | --- | --- | --- |
| MSE | 0.025 | 0.176 | 0.171 |
| MAE | 0.117 | 0.248 | 0.247 |
| RMSE | 0.158 | 0.420 | 0.413 |
| R2 | 0.993 | 0.951 | 0.951 |

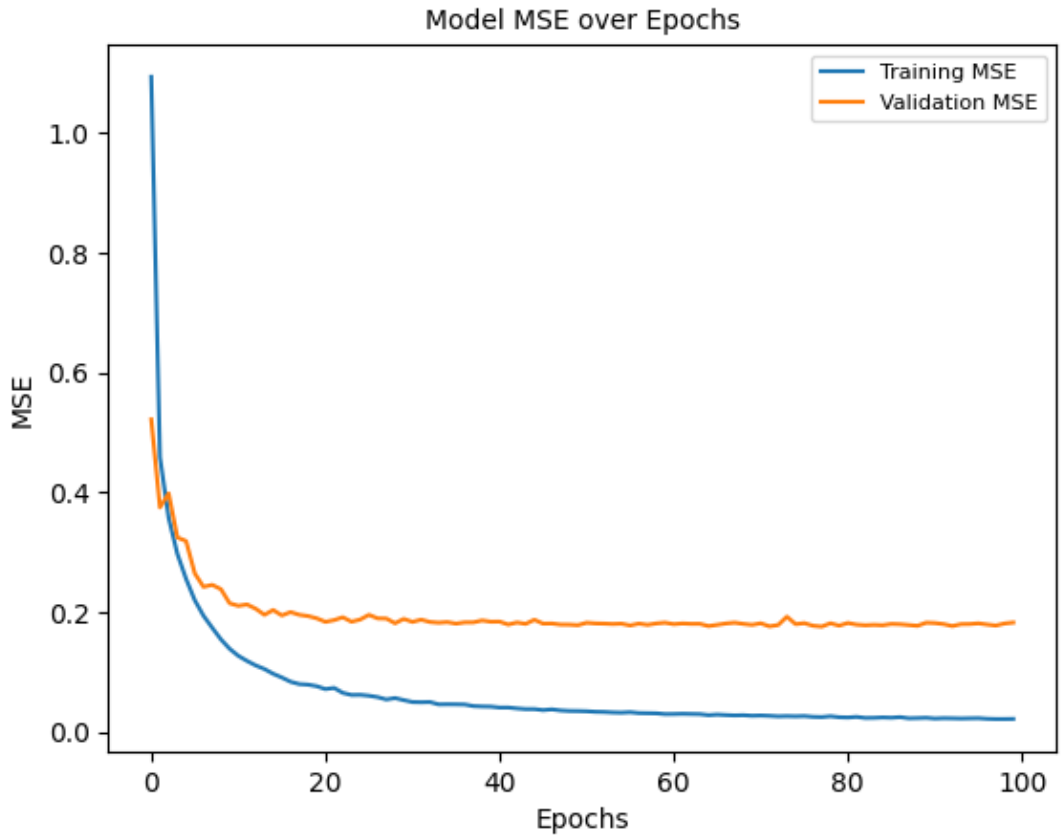

**Figure 3.** MSE of the model over training the model with 100 epochs

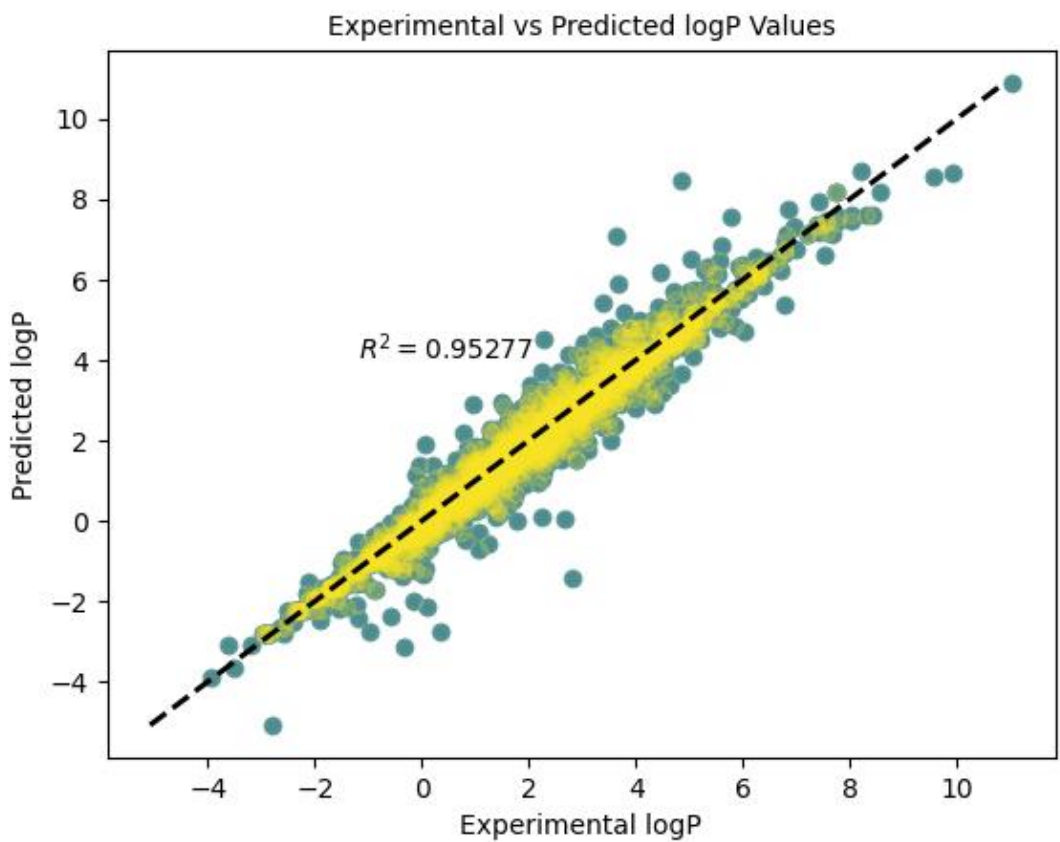

**Figure 4.** Correlation plot comparing predicted log P values with experimental values for compounds in the test set

**Testing the model on compounds with different functional groups.** To further evaluate the model's performance, one more time the dataset was divided into test, train, and validation sets. At first, 10% of the data was extracted as the test set, while the remaining data was used to train and validate the model. 20% of the data points were used for validation, and the remaining 80% was used to train the model. Additionally, to gain a perspective on the distribution and diversity of the test set, Q-Q plots and histograms were plotted, as shown in Figure 5. The distribution of the test set (Figure 5) was the same as the train set, with an almost normal distribution. The model was trained while monitoring the validation Mean Squared Error (MSE) and the model with the smallest MSE were saved as the best one. The best model was utilized to predict the log P for various categories of compounds, including different functional groups such as halogens, benzene rings, carboxylic acid, aldehyde, amide, aliphatic hydroxy, and aromatic hydroxy. According to Figure 6, the most accurate predictions were for compounds with aliphatic hydroxy groups, even more accurate than the predictions for the entire test set. Following the aliphatic hydroxy group, predictions were more accurate for chemicals with aromatic hydroxy and amide functional groups, with a negligible difference. Prediction accuracies for benzene and halogen, including chemicals, were nearly the same, with only a small difference of 0.003 log P between them. The lowest accuracy corresponded to the aldehydes, possibly due to the small number of compounds in the test (26) and train-validation set (186). The number of compounds with different functional groups and other metrics for accuracies are provided in Table S2.

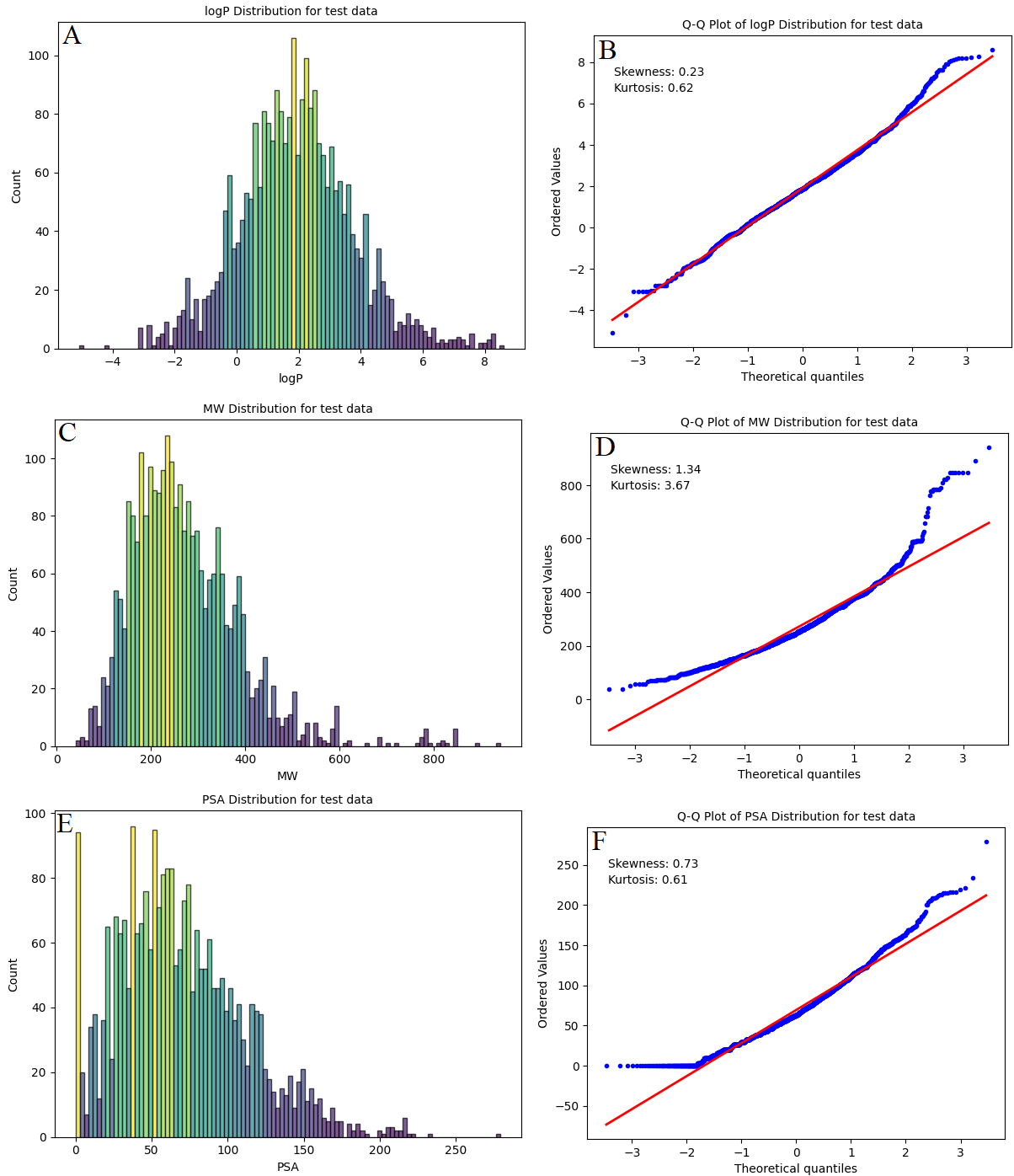

**Figure 5.** Histograms and Q-Q plots of MWs, PSA, and log P distributions of the test set

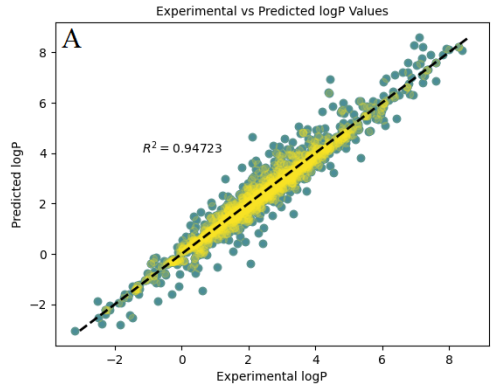
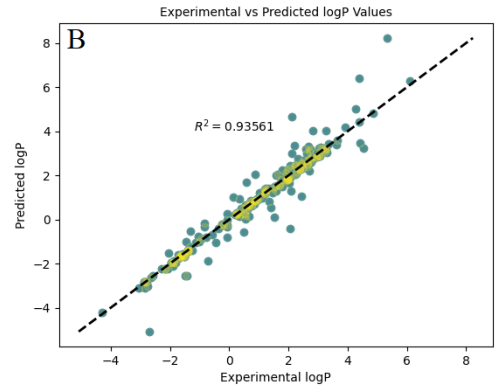
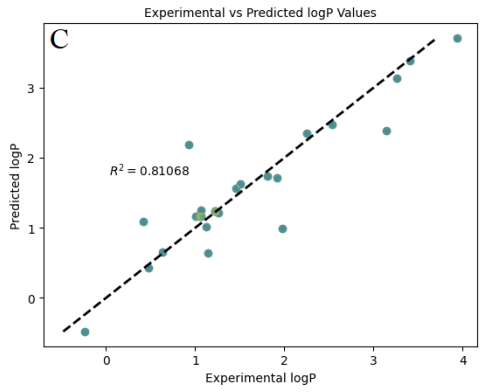
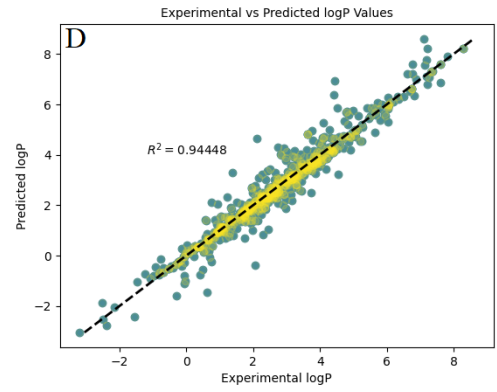
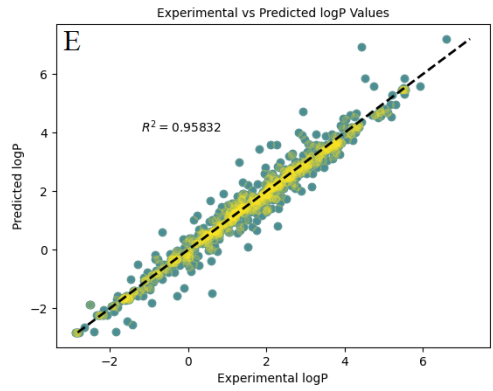
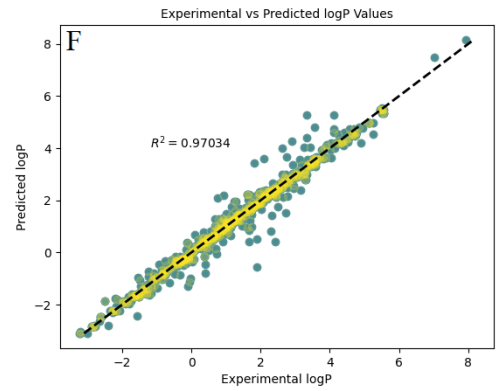
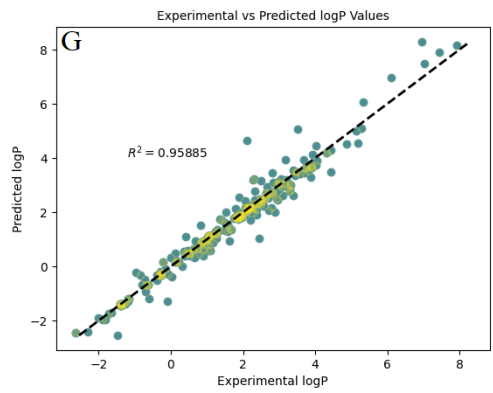

**Figure 6.** Correlation plots comparing predicted log P values with experimental values for compounds with A) Benzene Rings, B) Carboxylic Acid Functional Group, C) Aldehyde Functional Group, D) Halogens, E) Amide Functional Group, F) Aliphatic Hydroxy Groups, and G) Aromatic Hydroxy Groups

**Assessing the effect of dataset distribution on the model precision.** As illustrated in Figures 2A and 5A, the training and test datasets are saturated with compounds whose log P values are close to 2. Thus, the most reliable predictions were anticipated for the compounds with log P close to 2; however, as depicted by Figure 7B, the most accurate predictions were for compounds with log P values distributed around 4 and 2 (Figure 7B). If it were assumed that the accuracy level would be directly correlated with the number of instances with specific log P values in the training dataset, then this model should have provided the best results for the cases when the corresponding experimental log P values are close to 2. However, contrary to these assumptions, the highest prediction precisions were observed for the log P values of 2 and 4 (Figure 7B). In addition, the most unreliable estimations were observed for the log P values approximately equal to 1 (Figure 7A), even though there are more instances having log P around 1 than those having log P around 4 in both the training (Figure 2A) and test datasets (Figure 5A).

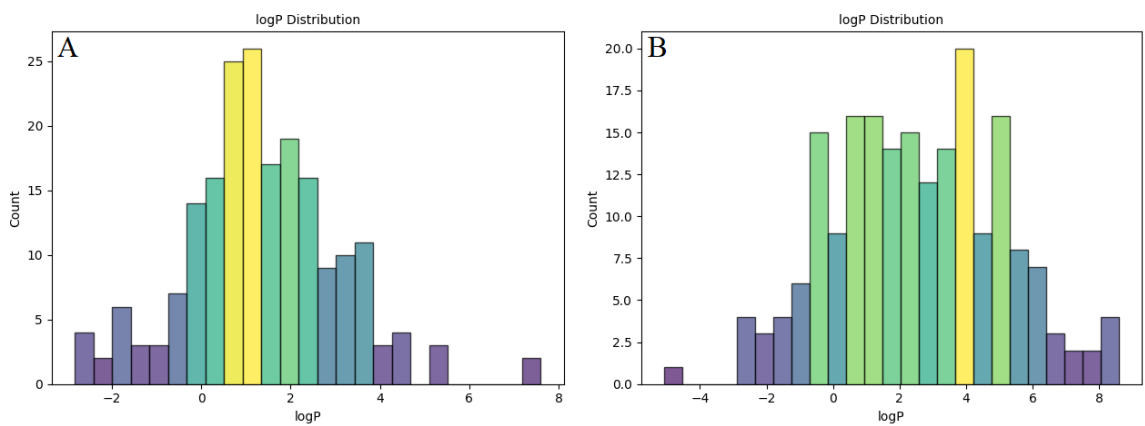

**Figure 7.** Experimental log P distributions for A) the least accurate predictions, and B) the most accurate predictions.

CONCLUSIONS

Overall, the development of the OWPCP deep-learning model is a significant step toward more precise predictions of the octanol-water partition coefficient, a crucial feature in many fields, particularly drug discovery. OWPCP effectively recognizes the essential structural patterns in a compound by utilizing Morgan fingerprints and MACCS keys. Therefore, making very accurate log P predictions without requiring experimental data obtained through painstaking experimental measurements. Strong training and validation of the model against an exhaustive dataset of compounds comprising 18,902 training, 4726 validation, and 2626 test instances demonstrated its excellence with a mean absolute error of 0.247, surpassing all the current computational methods and proving the strong predictive capability of the model across a diverse set of functional groups. OWCPC not only allows faster and more efficient prediction in the early stage of drug development but can also be applied to avoid problems with experimental measurement processes, including unstable or hardly synthesizable compounds. For further refinement, using larger and more complex datasets might not enhance the model's capability since there is inconsistency in the data distribution and metrics accuracy (more data on specific points did not enhance the model's accuracy for predictions within the related range). Hence, further expansion of the dataset is unlikely to result in better performance for the model. Instead, increasing the complexity of the model and further inclusion of additional descriptors can lead to better results.

ASSOCIATED CONTENT

**Supporting Information**.

A PDF file including Tuned Hyperparameters of the model, distribution of compounds with different functional groups in the test set, and accuracy metrics for each category in the test set is available, as well as A CSV file including SMILES, and experimental log p values used to develop the model.

AUTHOR INFORMATION

**Corresponding Author**

*Mohammadjavad Maleki - E-mail: Jmohammadmaleki@gmail.com

**Present Addresses**

†Department of Chemistry, University of Michigan, Ann Arbor, MI, USA

**Author Contributions**

M. Maleki conceptualized and developed the model. Both authors contributed to refining the initial concept, as well as to the writing and revision of the manuscript.

**Funding Sources**

This research did not receive any financial support or funding.

**Notes**

The authors declare no competing financial interest.

**Data and Software Availability**

In addition to the dataset, python scripts, and model described in this work, the trained model and a tutorial on how to use the trained model for log P prediction can be publicly accessed at the open-source GitHub repository "OWPCP" (https://github.com/jmohammadmaleki/OWPCP.git). The software tools used in this study, including RDKit (https://www.rdkit.org/), Matplotlib (https://matplotlib.org/), scikit-learn (https://www.scikit-learn.org/), NumPy (https://www.numpy.org/), Keras (https://keras.io/), TensorFlow (https://www.tensorflow.org/), Keras Tuner (https://keras.io/keras_tuner/), and Pandas (https://www.pandas.pydata.org), are freely available at their websites.

ACKNOWLEDGMENT

The authors would like to thank Google for providing the Google Colab platform, which facilitated the computational experiments and analyses in this work.

ABBREVIATIONS

MLP, multilayer perceptron; DNN, deep neural networks; ML, machine learning; RMSE, root mean-squared error; MSE, mean-squared error, MAE, mean absolute error; $R^2$, $R^2$-Score; SMILES, Simplified Molecular Input Line Entry System;

REFERENCES

[1]  S. Moldoveanu and V. David, "Characterization of analytes and matrices," in *Essentials in Modern HPLC Separations*, 2022. doi: 10.1016/b978-0-323-91177-1.00003-x.

[2]  R. N. Smith, C. Hansch, and M. M. Ames, "Selection of a reference partitioning system for drug design work," *J Pharm Sci*, vol. 64, no. 4, 1975, doi: 10.1002/jps.2600640405.

[3] S. Amézqueta, X. Subirats, E. Fuguet, M. Roses, and C. Rafols, "Octanol-water partition constant," in *Liquid-Phase Extraction*, 2019. doi: 10.1016/B978-0-12-816911-7.00006-2.

[4] C. A. Lipinski, "Lead- and drug-like compounds: The rule-of-five revolution," 2004. doi: 10.1016/j.ddtec.2004.11.007.

[5] C. A. Lipinski, F. Lombardo, B. W. Dominy, and P. J. Feeney, "Experimental and computational approaches to estimate solubility and permeability in drug discovery and development settings," 1997. doi: 10.1016/S0169-409X(96)00423-1.

[6] J. A. Arnott and S. L. Planey, "The influence of lipophilicity in drug discovery and design," 2012. doi: 10.1517/17460441.2012.714363.

[7] M. J. Waring, "Lipophilicity in drug discovery," 2010. doi: 10.1517/17460441003605098.

[8] A. Tsantili-Kakoulidou and V. J. Demopoulos, "Drug-like Properties and Fraction Lipophilicity Index as a combined metric," *ADMET DMPK*, vol. 9, no. 3, 2021, doi: 10.5599/admet.1022.

[9] J. Sangster, "Octanol Water Partition Coefficients of Simple Organic Compounds," *J Phys Chem Ref Data*, vol. 18, no. 3, 1989, doi: 10.1063/1.555833.

[10] T. Fujita, J. Iwasa, and C. Hansch, "A New Substituent Constant, ir, Derived from Partition Coefficients," *J Am Chem Soc*, vol. 86, no. 23, 1964, doi: 10.1021/ja01077a028.

[11] A. Leo, C. Hansch, and D. Elkins, "Partition coefficients and their uses," 1971. doi: 10.1021/cr60274a001.

[12] H. van de Waterbeemd, *Chemometric Methods in Molecular Design*, vol. 2. 2008. doi: 10.1002/9783527615452.

[13] A. Merritt, "Combinatorial Library Design and Evaluation: Principles, Software Tools and Applications in Drug Discovery," *Drug Discov Today*, vol. 7, no. 1, 2002, doi: 10.1016/s1359-6446(01)02103-1.

[14] Q. Liao, J. Yao, and S. Yuan, "SVM approach for predicting LogP," *Mol Divers*, vol. 10, no. 3, 2006, doi: 10.1007/s11030-006-9036-2.

[15] N. Ulrich, K. U. Goss, and A. Ebert, "Exploring the octanol–water partition coefficient dataset using deep learning techniques and data augmentation," *Commun Chem*, vol. 4, no. 1, 2021, doi: 10.1038/s42004-021-00528-9.

[16] S. Prasad and B. R. Brooks, "A deep learning approach for the blind logP prediction in SAMPL6 challenge," *J Comput Aided Mol Des*, vol. 34, no. 5, 2020, doi: 10.1007/s10822-020-00292-3.

[17] R. Lui, D. Guan, and S. Matthews, "A comparison of molecular representations for lipophilicity quantitative structure–property relationships with results from the SAMPL6 logP Prediction Challenge," *J Comput Aided Mol Des*, vol. 34, no. 5, 2020, doi: 10.1007/s10822-020-00279-0.

[18] K. Mansouri, C. M. Grulke, R. S. Judson, and A. J. Williams, "OPERA models for predicting physicochemical properties and environmental fate endpoints," *J Cheminform*, vol. 10, no. 1, 2018, doi: 10.1186/s13321-018-0263-1.

[19] R. Datta, D. Das, and S. Das, "Efficient lipophilicity prediction of molecules employing deep-learning models," *Chemometrics and Intelligent Laboratory Systems*, vol. 213, 2021, doi: 10.1016/j.chemolab.2021.104309.

[20] A. Yoshimori, "Prediction of molecular properties using molecular topographic map," *Molecules*, vol. 26, no. 15, 2021, doi: 10.3390/molecules26154475.

[21] E. B. Lenselink and P. F. W. Stouten, "Multitask machine learning models for predicting lipophilicity (logP) in the SAMPL7 challenge," *J Comput Aided Mol Des*, vol. 35, no. 8, 2021, doi: 10.1007/s10822-021-00405-6.

[22] Z. M. Win, A. M. Y. Cheong, and W. S. Hopkins, "Using Machine Learning To Predict Partition Coefficient (Log P) and Distribution Coefficient (Log D) with Molecular Descriptors and Liquid Chromatography Retention Time," *J Chem Inf Model*, vol. 63, no. 7, 2023, doi: 10.1021/acs.jcim.2c01373.

[23] M. Popova, O. Isayev, and A. Tropsha, "Deep reinforcement learning for de novo drug design," *Sci Adv*, vol. 4, no. 7, 2018, doi: 10.1126/sciadv.aap7885.

[24] K. Mansouri, C. M. Grulke, A. M. Richard, R. S. Judson, and A. J. Williams, "An automated curation procedure for addressing chemical errors and inconsistencies in public datasets used in QSAR modelling," *SAR QSAR Environ Res*, vol. 27, no. 11, 2016, doi: 10.1080/1062936X.2016.1253611.

[25] G. Landrum, "RDKit : A software suite for cheminformatics , computational chemistry , and predictive modeling," 2013.

[26]  D. Rogers and M. Hahn, "Extended-connectivity fingerprints," *J Chem Inf Model*, vol. 50, no. 5, 2010, doi: 10.1021/ci100050t.

[27]  J. L. Durant, B. A. Leland, D. R. Henry, and J. G. Nourse, "Reoptimization of MDL keys for use in drug discovery," *J Chem Inf Comput Sci*, vol. 42, no. 6, 2002, doi: 10.1021/ci010132r.

[28]  J. D. Hunter, "Matplotlib: A 2D graphics environment," *Comput Sci Eng*, vol. 9, no. 3, 2007, doi: 10.1109/MCSE.2007.55.

[29]  F. Chollet and others, "Keras," 2015.

[30]  M. Rucci and A. Casile, "{TensorFlow}: Large-Scale Machine Learning on Heterogeneous Systems}," Network: Computation in Neural Systems.

[31]  Python Software Foundation, "Python Programming Language," 2023. [Online]. Available: https://www.python.org/

[32]  D. E. Rumelhart, G. E. Hinton, and R. J. Williams, "Learning representations by back-propagating errors," *Nature*, vol. 323, no. 6088, 1986, doi: 10.1038/323533a0.

[33]  M. R. El Khili, S. A. Memon, and A. Emad, "MARSY: A multitask deep-learning framework for prediction of drug combination synergy scores," *Bioinformatics*, vol. 39, no. 4, 2023, doi: 10.1093/bioinformatics/btad177.

[34]  T. O'Malley, E. Bursztein, J. Long, F. Chollet, H. Jin, and L. Invernizzi, "Keras tuner," *Retrieved May*, vol. 21, 2019.

[35] L. Li, K. Jamieson, G. DeSalvo, A. Rostamizadeh, and A. Talwalkar, "Hyperband: A novel bandit-based approach to hyperparameter optimization," *Journal of Machine Learning Research*, vol. 18, 2018.

[36] X. Glorot, A. Bordes, and Y. Bengio, "Deep sparse rectifier neural networks," in *Journal of Machine Learning Research*, 2011.

[37] D. A. Clevert, T. Unterthiner, and S. Hochreiter, "Fast and accurate deep network learning by exponential linear units (ELUs)," in *4th International Conference on Learning Representations, ICLR 2016 - Conference Track Proceedings*, 2016.

[38] R. Barret, "Importance and Evaluation of the Polar Surface Area (PSA and TPSA)," in *Therapeutical Chemistry*, 2018. doi: 10.1016/b978-1-78548-288-5.50005-6.

[39] E. H. Kerns and L. Di, "Effects of Properties on Biological Assays," in *Drug-like Properties: Concepts, Structure Design and Methods*, 2008. doi: 10.1016/b978-012369520-8.50041-3.

[40] Z. M. Win, A. M. Y. Cheong, and W. S. Hopkins, "Using Machine Learning To Predict Partition Coefficient (Log P) and Distribution Coefficient (Log D) with Molecular Descriptors and Liquid Chromatography Retention Time," *J Chem Inf Model*, vol. 63, no. 7, 2023, doi: 10.1021/acs.jcim.2c01373.

# Supporting Information:

# OWPCP: A Deep Learning Model to Predict Octanol-Water Partition Coefficient

*Mohammadjavad Maleki*[*,1]*, Sobhan Zahiri†[2]*

1. Department of Chemistry, Sharif University of Technology, Azadi Ave, Tehran, Iran

2. Department of Chemistry, Isfahan University of Technology, University Boulevard, Esteghlal Square, Isfahan, Iran

This document contains Tables S1, and S2

## TABLE OF CONTENTS

**Table S1.** Hyperparameters of the model architecture

| Encoder | Layer | Hyperparameter | |
|---|---|---|---|
| MACCS | 1 (input) | Units | 166 |
| | 2 (hidden) | Units | 2318 |
| | | Activation Function | Tanh |
| | | Dropout | 0.3 |
| | 3 (output) | Units | 1802 |
| | | Activation Function | ReLU |
| MFP | 1 (input) | Units | 2048 |
| | 2 (hidden) | Units | 2834 |
| | | Activation Function | ELU |
| | | Dropout | 0.1 |
| | 3 (output) | Units | 2834 |
| | | Activation Function | Linear |
| Log P predictor | 1 (input) | Units | 1802 |
| | | Activation Function | ReLU |
| | 2 (hidden) | Units | 2576 |
| | | Activation Function | Tanh |
| | | Dropout | 0.25 |
| | 3 (output) | Units | 1 |
| | | Activation Function | Linear |

**Table S2.** Distribution of compounds with different functional groups in test set, and accuracy metrics for each category

| Functional Group | Number of instances within the testset | Metrics | |
|---|---|---|---|
| Benzene | 1707 | MSE | 0.154 |
| | | MAE | 0.245 |
| | | RMSE | 0.392 |
| | | R2 | 0.947 |
| Carboxylic acid | 216 | MSE | 0.282 |
| | | MAE | 0.285 |
| | | RMSE | 0.531 |
| | | R2 | 0.936 |
| Aldehyde | 25 | MSE | 0.166 |
| | | MAE | 0.251 |
| | | RMSE | 0.407 |
| | | R2 | 0.811 |
| Halogen | 681 | MSE | 0.191 |
| | | MAE | 0.267 |
| | | RMSE | 0.437 |
| | | R2 | 0.944 |
| Amide | 741 | MSE | 0.134 |
| | | MAE | 0.228 |
| | | RMSE | 0.366 |
| | | R2 | 0.958 |
| Aliphatic OH | 654 | MSE | 0.112 |
| | | MAE | 0.184 |
| | | RMSE | 0.334 |
| | | R2 | 0.970 |

| Aromatic OH | 239 | MSE | 0.146 |
|---|---|---|---|
| | | MAE | 0.229 |
| | | RMSE | 0.381 |
| | | R2 | 0.959 |